\pdfoutput=1
\documentclass[letterpaper]{article}
\usepackage{aaai}
\usepackage{times}
\usepackage{helvet}
\usepackage{courier}
\usepackage{graphicx} 
\usepackage{url} 
\usepackage{makecell} 
\usepackage{xcolor} 
\definecolor{mygreen}{rgb}{0, 0.6, 0} 

\begin{document}

\title{A Mental Trespass? Unveiling Truth, Exposing Thoughts and Threatening Civil Liberties with Non-Invasive AI Lie Detection}

\author{AAAI Press\\
AAAI\\
2275 East Bayshore Road, Suite 160\\
Palo Alto, California 94303\\
}

\author{Taylan Sen, Kurtis Haut, Denis Lomakin, Ehsan Hoque\\
tsen@cs.rochester.edu, khaut@u.rochester.edu, dlomaki2@u.rochester.edu, mehoque@cs.rochester.edu\\
University of Rochester\\
Deptment of Computer Science\\
}

\maketitle
\begin{abstract}
Imagine an app on your phone or computer that can tell if you are being dishonest, just by processing affective features of your facial expressions, body movements, and voice. People could ask about your political preferences, your sexual orientation, and immediately determine which of your responses are honest and which are not. In this paper we argue why artificial intelligence-based, non-invasive lie detection technologies are likely to experience a rapid advancement in the coming years, and that it would be irresponsible to wait any longer before discussing its implications. Legal and popular perspectives are reviewed to evaluate the potential for these technologies to cause societal harm. To understand the perspective of a “reasonable” person, we conducted a survey of 129 individuals, and identified consent and accuracy as the major factors in their decision-making process regarding the use of these technologies.  In our analysis, we distinguish two types of lie detection technology: ``accurate truth metering" and ``accurate thought exposing". We generally find that truth metering is already largely within the scope of existing US federal and state laws, albeit with some notable exceptions. In contrast, we find that current regulation of thought exposing technologies is ambiguous and inadequate to safeguard civil liberties. In order to rectify these shortcomings, we introduce the legal concept of ``mental trespass" and use this concept as the basis for proposed regulation.
\end{abstract}

\section{Introduction}

\textit{``Beyond my expectation, thru uncontrollable factors, this scientific investigation became for practical purposes a Frankenstein’s monster, which I have spent over 40 years in combating."}

\vspace{0.5cm}

These are the words of the first U.S. policeman with a PhD in science, John Larson, reflecting on his invention of the contemporary polygraph shortly before his death \cite{alder2007lie}. Larson was troubled by the improper use of and the unreasonable level of trust placed in the polygraph and the harm caused by the many who were unfairly accused of dishonesty. Over the years since the first practical application of the polygraph in 1921, numerous job applicants were denied employment and government employees lost their jobs \cite{grubin2005lie}. Over two million Americans were being tested each year by the 1980s \cite{goldzband1990polygraph}. It was not until the introduction of the Employee Polygraph Protection Act in 1988 that use of the polygraph and similar devices was banned from most employment settings. While it took the U.S. legislature 67 years to formally regulate the polygraph, the Federal judicial system barred polygraph-like devices in their first application in the courtroom in 1923 \cite{Frye1923}. The court in \textit{Frye v. United States} established that in order to ``admit expert testimony deduced from a well-recognized scientific principle or discovery, the thing from which the deduction is made must be sufficiently established to have gained general acceptance in the particular field in which it belongs". The D.C. Circuit court denied defendant Alfonso Frye's attempt to use the \textit{blood pressure deception test} (the precursor to Larson's polygraph based on periodic blood pressure readings alone) administered by Harvard psychologist Dr. William Marston to establish his innocence to the murder charges he faced. The ``Frye standard" was rapidly adopted by most states' judicial systems \cite{jensen2002frye}. In later analysis of this paper, we show that a legal dichotomy still exists with regards areas of the law which are well defined and others which are largely ambiguous with regards to lie detection technologies, and that this ambiguity is likely to cause harm.   

Larson's quote in mentioning the story of Dr. Frankenstein's creation of an artificially intelligent creature which becomes a murderous monster is particularly prescient in light of the recent advances AI systems have made in recent years. While we benefit from ever more surprising contributions to our daily lives - AI powered thermal cameras systems are actively being used to screen passengers for fevers associated with coronavirus \cite{Hochreutiner2020}, systems can even extract heart rate from common video stream or wifi signals for health monitoring \cite{wang2018comparative}, \cite{lee2018design}; AI systems, like Frankenstein's monster, also bring the potential to cause substantial harm. With the increasing powers of noninvasive AI also come new methods for invasion of privacy and circumvention of our rights against unreasonable search. For example, a recent AI system purports to be able to predict one’s sexual orientation from their facial features \cite{wang2018deep}. It is easy foresee the harm that can result from exposing one’s private sexuality considering the case of the Rutgers University student who died of suicide after his roommate set up a webcam and publicly broadcast a private homosexual encounter with another student \cite{Pilkington2010}. Similarly daunting is the recent growth in use of AI systems in police surveillance. The Chinese government has been accused of oppression against the Uighgur minorities in the Xinjian province as AI facial recognition systems have been extensively deployed \cite{bbc2020}. Chinese authorities state that use of such technologies are necessary to fight terrorism and that similar surveillance systems were instrumental in enforcing the quarantines that helped halt the progression of coronavirus throughout China \cite{fong2020artificial}. How can we fix ambiguities in the law to allow us to benefit from AI sensing technology advances while ensuring that they are not abused?

Reynolds and Picard were one of the first to examine the related question of ``Would  it  be  ethical  for  a  computer  to  sense  a  user’s  emotions?"  \cite{reynolds2004affective}. Through an online survey (N=125) they found that respondents were more likely to consider a system that collects and exposes one's emotions as ethical if there is an explicit contract for which users consent. Calls have been made for ``design contractualism" in which systems which are capable of reading user's emotions are designed to obtain consent \cite{pitt2012design}. In addition to supporting this contractual premise, Reynolds and Picard also promote an analysis of the social dimensions of how a product will be used in determining whether such an application is ethical \cite{reynolds2005evaluation}, \cite{reynolds2004affective}. In situations where consent is not provided, analysis of AI sensing systems in a civil liberties context is largely limited to facial recognition \cite{raji2020saving},\cite{brey2004ethical} and have even been the focus of a recent congressional oversight committee hearing \cite{oversight2019}.

In this paper we examine the progression of lie detection technologies and consider to what extent US law currently regulates coming technologies. To aid our analysis, we define two types of lie detection technology: \textit{truth metering}, which involves using a device evaluating one's level of belief in one of their statements, and \textit{thought exposing}, which involves using a device to predict an individual's inner thoughts.

In summary, we generally find that truth metering is already largely within the scope of existing US federal and state laws, albeit with some notable exceptions. In contrast, we find that current regulation of thought exposing technologies is ambiguous and inadequate to safeguard privacy and civil liberties. In order to rectify these shortcomings, we introduce the legal concept of ``mental trespass" and use this concept as the basis for proposed legislation.

More specifically, in this paper we argue that:
\begin{itemize}
\item Development of non-invasive, AI-based lie detection technologies are likely to progress rapidly in the near future, and no law or government effort is likely to halt its production, distribution, and use (in many cases the government is investing heavily in the advancement of such technologies).
\item Lie detection technologies carry with them much potential for individual harm in terms of loss of privacy, wrongful criminal conviction, and unfair bias.
\item While the current legal environment generally already regulates truth metering technologies, it is largely ambiguous with regards to the legality of uses of thought exposing technologies.
\item In order to mitigate the potential harms such technologies may bring, we recommend the introduction of a regulatory federal ``Mental Trespass Act", which calls for a general use ban of non-consensual use of thought exposing technology, but allows for non-offensive uses of truth metering devices. 
\end{itemize}

\section{Technology Progression: A Lie Detection Revolution}
Lie detection was essential enough to human civilizations that it appears in the Code of Hammurabi, one of the very first instances of written law from circa 1754 BC \cite{roth1995mesopotamian}. Translations of preserved tablets of the Code state that questions of honesty were to be resolved through what has been termed \textit{trial by ordeal}: ``If anyone  bring  an  accusation  against a  man, and the accused  ... jump  into  the  river  ...  if  he  sink  in  the  river  his accuser shall take possession of his house..." \cite{roth1995mesopotamian}. What started out as random change and religions belief, the state of the art in lie detection and related law has progressed to include increasingly powerful scientific techniques including advanced sensing tools and more refined questioning techniques as depicted in the Timeline in Fig. 2. It took approximately 800 more years after Hammurabi before the first glimmer of scientific legitimacy in lie detection to appear, which was found in the ancient Hindu text: the Vedas. Loosely based on the involuntary fight or flight response (which causes individuals to go white as blood is diverted from body extremities to the heart and lungs), the Vedas describes how to spot a murderer, \textit{``[The poisoner] … does not answer questions, or they are evasive answers; he speaks nonsense … his face is discolored …"} \cite{Trovillo1939}. The scientific progression of lie detection continued in the 3rd century BC, as renowned physician Erasistratus used pulse, skin temperature, and skin pallor, to correctly detect the lies of Prince Antiochus, as the prince tried to conceal his passionate love for his father's new wife \cite{trovillo1938history}, \cite{amsel2019planting}.

An underlying premise regarding lie detection began to be recognized. Charles Darwin wrote in his 1872 book, \textit{The Expression of Emotions in Man and Animals}, that ``...actions become habitual in association with certain states of the mind, and are performed whether or not of service in each particular case..." \cite{Darwin1873}. We recognize a fundamental premise of lie detection in that \textit{a person's internal state of mind uncontrollably leaks out into the externally observable world when appropriately probed}. Indeed, this premise must hold for a given lie detection technique to work. Through \textit{appropriately probed} we recognize that specialized questioning techniques may be necessary to cause honest and dishonest subjects to elicit detectable differences in observable behavior. This definition additionally brings attention that advanced tools may be useful in observing these subtle differences. 

\begin{figure}[htbp]
\centerline{\includegraphics[width=1\linewidth]{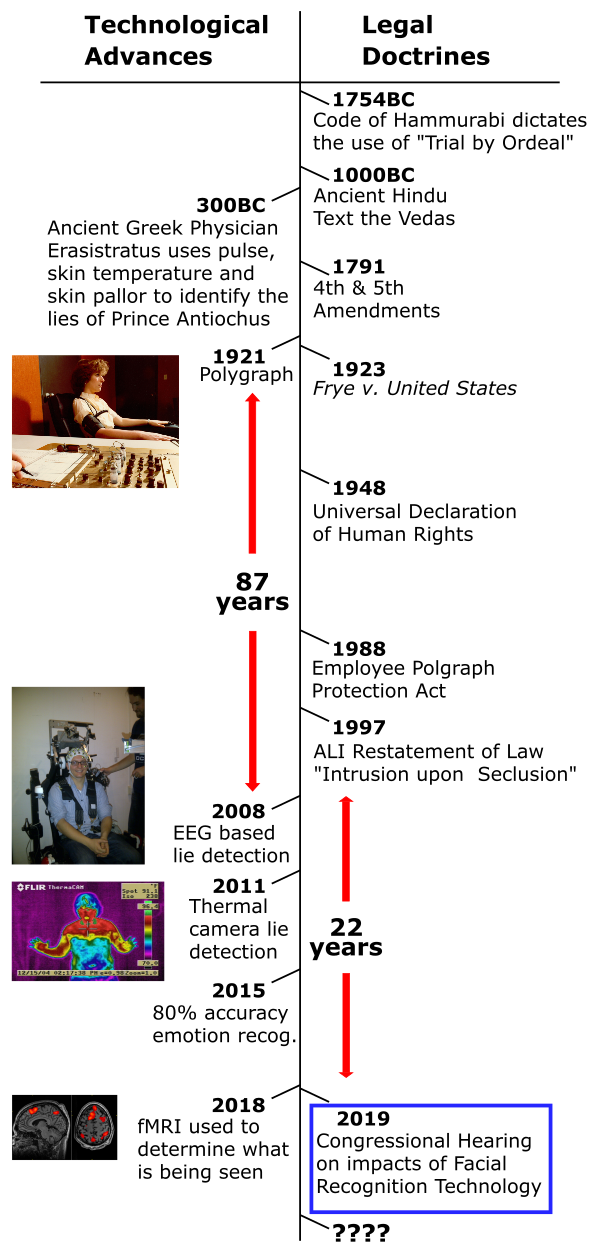}}
\caption{\textbf{Side by Side Timeline of Technological Advances and Legal Doctrines} 
(Note: not to true exponential scale)}
\label{fig_timeline}
\end{figure} 

The use of tools and specialized questioning techniques in lie detection is demonstrated with the perhaps most well-known and widely used lie detection device, the contemporary polygraph. Like Erasistratus's technique, the common polygraph tracks the subject's heart rate and respiration. The modern polygraph, however, has two notable improvements over Erastratus including: 1) additional sensors for blood pressure, skin conductivity, and respiration rate; and 2) a formal questioning technique, known as the \textit{control question test}. The polygraph sensors collectively provide a measure of the subject's physiological arousal. Crucially, the control question test begins with questions unrelated to the matter for which the lie detector is being applied, including \textit{baseline questions} and \textit{control questions}. Baseline questions are trivial questions used to indicate the subject's arousal at rest. Alternatively, control questions are designed to create a strong physiological response in most people, for example  \textit{Have you ever stolen office supplies from work? Have you ever cheated on your taxes?}. The unrelated questions are followed with \textit{relevant questions}, which are pertinent to whatever is being investigated. The underling theory of the control question test is that someone who is lying is more likely to be nervous during the relevant questions than during the control questions, compared to an honest subject who is expected to have a stronger level of arousal during the control questions \cite{raskin2002comparison} \cite{bradley1987machiavellianism}. Other questioning techniques such as the guilty knowledge test (GKT), which relies on strategic use of information only a guilty person would have, have been developed and compared with the control question test \cite{myers1997polygraph}. The GKT has been particularly used in systems which rely on advanced sensing, including electroencephalograms (EEG) and function MRI (fMRI). In addition to the advances in questioning techniques and sensors, perhaps most revolutionary recently has the advance in data analysis: including both raw computing power as well as advanced machine learning techniques. 

The world's first real supercomputer was Control Data Corporation's CDC 6600, developed in 1964 \cite{Hosch2018}. The computer was enormous, the size of multiple people, and state of the art - miles ahead of the competition. Three times as fast as its predecessor, it could run 3 million megaFLOPS. It cost the equivalent of \$60 million in 2018 \cite{Spicer2000,BLS}. The CDC 6600 was so powerful the word ``supercomputer" was coined to describe it. If someone were to tell its creator, Seymour Cray, that in 50 years' time a processor the size of his forearm would cost 50,000 times less and be 2 million times faster, he might not believe them. But the NVIDIA GeForce GTX Titan X, released in 2015, was exactly that \cite{NVIDIA2018,Hruska2016}.

Piggybacking off of the recent hardware advances, several algorithmic successes have been made in the field of computer vision. In 2012, ``AlexNet" astounded researchers with its accuracy in image classification and demonstrated the power of convolutional neural networks for the task\cite{krizhevsky2012imagenet}. In 2014, the invention of generative adversarial networks utilized deep learning to generate realistic images \cite{goodfellow2014generative}, which recently became embroiled in controversy with their application in deepfakes. Researchers and software engineers working with computer vision have an incredible array of tools with which to develop new technologies in the coming years. We highlight the progress in computer vision specifically because these advances enable lie detection to be performed at a distance due to their inherent non-invasiveness.

Additionally, one of the major factors limiting progress in deception detection is the lack of good data. However, with recent advances in Internet technologies, techniques are available to gather data on deception. For example, Sen et al. developed a system for gathering video deception data via crowdsourced individuals\cite{sen2018automated}. In addition, US government entities have very recently expressed desire to gather data sets on ``credibility assessment", which could be used to develop deception detection technologies. In fact, during 2019, the Intelligence Advanced Research Project Association (IARPA) put out a grand challenge concerning the collection of deception data. The Credibility Assessments Standardized Evaluation (CASE) Challenge formally called for a protocol to standardize this procedure in regards to how these datasets are gathered and accessed. Additionally, governments have started pouring vast amounts of funding into projects which expand their powers of surveillance. Backed by the Chinese and Russian governments, AI startup Megvii raised \$460 million for the development of facial recognition technology. 

We emphasize these specific examples to illustrate the non-invasive nature of these developing technologies, advancements in data collection procedures/capabilities and government interest. Because of these qualities, it is inevitable that accurate AI-based lie detection will soon be upon us. 

\section{Laws and Limitations: Current US Federal and State Laws}
We next analyze current laws regarding the public use of non-invasive deception detection technology without an observed party's consent. While our focus is on U.S. law, it is worth noting that the 1948 Universal Declaration of Human Rights has explicit language regarding human rights to ``privacy". More specifically, Article 12 of the declaration states ``No one shall be subjected to arbitrary interference with his privacy, family, home or correspondence...". While the interpretation of what constitutes ``arbitrary interference" and ``privacy" are left undefined, it is noteworthy that any notion of such a privacy right was considered so important as to be codified in the Universal Declaration.

\subsection{4th Amendment}
In the United States, perhaps the most relevant legal issue with regards to public deception detection is raised with regards to the 4th amendment. The 4th amendment establishes the ``right of the people to be secure in their persons...against unreasonable searches" and has been interpreted to prohibit searches when there is a reasonable expectation of privacy \cite{amar1994fourth}. Several cases have established that in general there is no reasonable expectation of privacy for things which are in plain view in a public area.  For example, the U.S. Supreme Court held that garbage that is left out on the curb can be searched without a warrant in California v. Greenwood \cite{herdrich1988california}\cite{simpson1989california}\cite{cunis1988california}. This has been extended to include use of some specialized equipment, particularly the use of a plane for aerial observation of someone's backyard in California v. Ciraolo \cite{falcone1986california}, and observation of an open field in an industrial complex with a high definition camera in Dow Chemical Co. v. United States \cite{joyce1982epa}\cite{ruzi1988reviving}. The court seemed to indicate the relevance of whether the equipment was available to the public, in one case finding that the EPA did not violate the 4th amendment when it “was not employing some unique sensory device not available to the public".  An analysis of the smells during a routine traffic stop with a specialized drug-sniffing dog was also found to not constitute an unreasonable search in Illinois v. Caballes \cite{dery2005let}\cite{simmons2005two}\cite{bekiares2005constitutional}. However, the ability to observe someone from a public area is not absolute. The Supreme Court found in Kyllo v. United States \cite{seamon2001kyllo}\cite{adkins2001supreme}\cite{froh2002rethinking} that viewing a person's home from outside with a thermal imaging camera (to determine if high temperature drug growing lights were used) was indeed a violation of one's ``reasonable expectation of privacy".
In light of these Supreme Court cases regarding 4th amendment rights, how would we expect the use of a video-based lie detection apparatus to play out? One perspective is that an individual's facial expressions are in plain view and thus do not carry a reasonable expectation of privacy, as in California v. Ciraolo regarding a person's backyard, or someone's garbage on the curb in California v. Greenwood. It is likely that the camera used for deception detection need not be more advanced than the high resolution camera deemed to be acceptable in Dow Chemical v. United States. 
However, lie detection does involve use of state of the art AI-driven algorithms and computer vision techniques. It seems conceivable that a court could find such algorithms as uncovering someone's internal state in an invasive way. Additionally, we may expect a court to consider, as it did in Dow Chemical Co. v. United States, whether the equipment used is publicly accessible. Thus, whether such lie detection technology is made public or not will possibly affect whether its use constitutes a 4th amendment search (e.g., if it is made available to the public, the Government would not be using ``specialized technology"). However, it should also be noted that 4th amendment issues are limited to the government (or people working on behalf of the government) and does not apply to public at large.   

\subsection{5th Amendment}
In addition to the potential 4th amendment issues, the use of lie detection technology in a court of law by the prosecution may bring up Constitutional Law issues with regards to the 5th amendment protection against self-incrimination.  The 5th amendment provides that ``[n]o person shall be ... compelled ... to be a witness against himself"\cite{amar1995fifth} \cite{rubin2003square}. We foresee that use of a lie detection technology without a subject's consent may be interpreted as compelled testimony. However, the courts have interpreted the 5th amendment narrowly, giving the prosecution the right to compel the accused to provide a password to encrypted computer data \cite{wachtel2013give}\cite{cauthen2017fifth}. Additionally, the courts have determined that a suspect may be compelled to produce fingerprints, blood, and fingernail scrapings without violating the 5th amendment\cite{inbau1998self}\cite{inbau1998self}. Further, courts have even found that compelling a witness to provide a voice sample for identification does not trigger 5th amendment protections\cite{weintraub1956voice}. Thus, we believe that it is unlikely that a court would find use of an AI-driven lie detection technology to be a violation of one's 5th amendment rights. However, in certain contexts perhaps this is not the case. For example, Thompson argues that highly invasive lie detection technology, such as unconsented application of the fMRI, is likely to violate the 5th amendment due process law as it ``shocks the conscience" \cite{thompson2007brave}.  Therefore, we take the stance that the degree of invasiveness is what fundamentally defines this question of violating the 5th amendment. We bring to light in this paper that highly accurate non-invasive lie detection technologies are not only imminent, but their risk for infringing upon our civil liberties is much greater. This is due to the fact that non-invasive lie detection devices are able to lie detect non-consenting individuals from a distance. Furthermore, it is not clear whether these noninvasive methods would shock one's conscience enough to violate the 5th using Thompson's terminology.

\subsection{Employee Polygraph Protection Act}
The Employee Polygraph Protection Act (EPPA) prevents private employers from requiring job applicants or current employees to submit to a lie detector of any kind, but allows polygraphs to be used by certain sectors, namely government and security positions. However, according to its website, the fMRI based lie detection company No Lie MRI ``measures the central nervous system directly and such is not subject to restriction by these laws". As noted by Greely and Illes, the language used in the provision of the legislation is broad enough that loopholes like this are possible \cite{greely2007neuroscience}. Without an explicit amendment or judicial review, No Lie MRI could continue to offer its services to employers, violating the intention of the EPPA, but not the text of the law. The EPPA is limited to employer-employee relationships, and is silent with regards to public use.

\subsection{Invasion of Privacy Laws}
The strongest limitations on the public use of a non-invasive lie detection technology appear to arise from state law. While it is difficult to analyze each state's laws individually, a concise restatement of the preferred rules used by a majority of the states is available in the ``Restatement of Law", written by the American Law Institute (ALI). The Restatement provides the law of Intrusion Upon Seclusion, commonly referred to as ``invasion of privacy", which makes liable one who ``intentionally intrudes, physically or otherwise, upon the solitude or seclusion of another or his private affairs or concerns...if the intrusion would be highly offensive to a reasonable person". This liability extends even when there is no publication or use of the information obtained in violation. In general, surveillance from a public place is not intrusion upon seclusion, however, exceptions to this rule exist. In summary, it is unclear if when accurate noninvasive lie detection arrives, it will be legal to use on non-consenting individuals caught unawares.

\subsection{Court System}
In the federal court system, it is unclear whether even a highly accurate lie detector would be admitted as evidence. Currently, polygraph tests and their results are almost entirely inadmissible in a federal court under evidentiary rules. Polygraph results are what is known as ``highly prejudicial," meaning that regardless of the test's accuracy or even its relevance to the case at hand, hearing about it will bias the jury. If the polygraph indicates that the defendant has lied, despite its questionable accuracy, a jury may treat that as definitive proof that the defendant lied. Additionally, if they believe the defendant lied about material facts related to the case, that may indicate proof of guilt to a jury, no matter how relevant or irrelevant those facts are to the defendant's guilt or innocence. For these reasons, it is possible that even a 99\% accurate lie detector could be excluded from evidence, due to a judge fearing the jury will treat it as 100\% accurate.

There are currently two standards by which scientific evidence can gain admission into the courtroom depending upon jurisdiction, known as the Frye standard (introduced in the Introduction) and the Daubert standard. The Frye standard provides that in order to be admitted, the scientific basis for the evidence ``must be sufficiently established to have gained general acceptance in the particular field in which it belongs". Per this standard, computer AI-based lie detection is almost always not admitted in its current state, as the underlying technology is still developing and the accuracy of this method of lie detection has not been well established.

The Daubert standard, which has largely superseded the Frye standard in both federal court and most state courts, sets stricter guidelines for evidence being admitted, but leaves the decision up to the court rather than the scientific community. This includes the Frye standard of general acceptance as well as whether the scientific evidence has been tested, whether it has been peer reviewed, and whether it has a high rate of error. AI-based lie detection would also likely be kept out of courtrooms according to this standard due to its current lack of widespread testing and peer review. These two standards ensure that the courts are well equipped to keep potentially inaccurate scientific evidence out of the courtroom.

Whether or not a technology is admitted into the courtroom is of the utmost importance for the following reason. Historical review shows that once a technology is deemed as legitimate (e.g. fingerprint analysis) or as questionable (such as the polygraph), such characterizations are unlikely to be changed \cite{thompson2007brave}. A major issue with the polygraph was raised in case law, with the Oregon Supreme Ct. finding ``the use of the polygraph ha[s] the potential to dehumanize parties and witnesses, treating them or as ‘electrochemical systems to be certified as truthful or mendacious by a machine.’" \cite{peterson1989bud}. The Daubert standard, similar to the Frye standard has almost entirely prevented polygraph admissibility \cite{mccall1996misconceptions}. Given that recent legal analyses argue that fMRI should not be allowed at this time \cite{langleben2013using}, \cite{kittay2006admissibility} \cite{wolpe2005emerging}\cite{moreno2009future}, it is likely the Daubert standard will keep these coming technologies out of the courtrooms as well for the time being. Scholars have gone on to argue for the urgent need to regulate developing lie detection technologies, such as the neuroscience-based technologies upon which fMRI lie detection \cite{greely2007neuroscience}, \cite{moreno2009future}. 

Prior analysis has also come to the conclusion that looking at technologies like fMRI through analogy with blood test and/or forced testimony is inappropriate, arguing that ``the implicit assumption of mind-body dualism, which underlies this thinking, is dated and, most likely, no longer tenable" \cite{thompson2007brave}.  Scholars have argued the importance of considering legal implications of an advancing technology before it becomes ubiquitous, with Thompson stating ``if the existing scientific literature is indeed a harbinger of an important new technology, it will be to society's benefit that some thought have been put into its implications before its wide scale deployment."\cite{thompson2007brave} . All in all, the topic of advanced lie detection has received recent attention in ethical and legal contexts \cite{langleben2013using}, \cite{farah2014functional}, \cite{greely2007neuroscience}, \cite{tennison2012neuroscience}, \cite{moreno2009future}. However, precise definitions of the technologies in question and proposed legal doctrines offering a solution have yet to be fleshed with enough granularity. With the legal doctrine and case history being classified as ambiguous at best, there is a clearly a strong societal need to formally define what should be allowed regarding these evolving technologies. 

\section{Public Perspective}
To understand public opinion on the usage of these lie detection technologies, we sampled the population by conducting a survey using the crowdsourcing platform Amazon Mechanical Turk. We included demographic information on the survey and based on the responses, launched multiple surveys with participation requirements to ensure that the demographic distribution of our respondents resembled the demographic distribution of the United States. (See appendices for side by side comparisons of race, gender, political affiliation, education level and age.) We believe matching the distribution is essential for obtaining not only a reliable sample, but one where we can reasonably extrapolate to claim any kind of generalizable opinion.  We monitored the quality of the survey responses by implementing a control question and eliminated all survey responses where the desired condition did not hold (e.g. ``answer strongly agree to this question"). We also incorporated a free response question to our survey and removed responses where the length of the response was less than 10 characters in length. After all unsatisfactory data points were removed, we were left with N = 129 quality responses. We set out to investigate whether public opinion was in favor of or opposed to the legality of using these technologies on an individual without first obtaining their consent (crucially important for noninvasive lie detection technologies as they can be used on an individual caught blissfully unaware). The results from the survey responses were strongly indicative of opposition to unconsented usage.

We combined the agree and strongly agree responses to find the number in favor as well as the disagree and strongly disagree responses to find the number opposed. 33 people were in favor, 69 people opposed and 27 were neither for nor against this usage case. We conducted a proportions statistical test with our null hypothesis being that there is no difference in public opinion regarding this question (e.g. the number in favor is equal to the number opposed). After running the statistical test, the probability of that null hypothesis given our data was p=0.0001 or 1/100th of a percent (0.01\%). We thus reject the null hypothesis of there being no difference in public opinion and accept the alternative hypothesis that likely there is a difference (meaning a majority of people are opposed to unconsented use of these lie detection technologies). Based on public opinion, there is certainly concern over unregulated lie detection technology being used maliciously and we have an obligation as a society to mitigate that outcome. We are hopeful that our proposed ``Mental Trespass Act" and recommendations for updating the language in the EPPA to reflect our technology definitions would greatly aid this communal effort.

\section{Recommendations}

\subsection{Definitions and Proposed Law}
In providing legal recommendations on how to mitigate the potential harms and ambiguity in the field, we first define two types of relevant technology as well as the different categories of their usage. We distinguish two major classes of lie detection tools including (1) accurate truth metering, and (2) accurate thought exposing (Depicted in figure 1).  

Accurate truth metering is defined as \textit{Use of a device to measure an individual’s level of belief in an intentional statement made by the individual, with the device usage having an accuracy exceeding typical human performance.} A statement broadly includes spoken utterances, written text and drawings, bodily gestures, and other forms of communication. An intentional statement, requires that the statement maker has the mental intent to make the statement. Thus, a spontaneous gasp of surprise, or the unconscious blushing after hearing a question are not intentional statements. In defining accurate, we use an \textit{excedat-hominem} standard, i.e. a level of accuracy which clearly exceeds typical human ability. Thus, in defining an accurate truth meter, we consider the numerous studies on human performance regarding lie detection and note that this level of accuracy has been found to be approximately 54 \% \cite{bond2006accuracy}, even amongst expert judges \cite{bond2008individual}.

\begin{figure}[tbph]
\centerline{\includegraphics[width=1\linewidth]{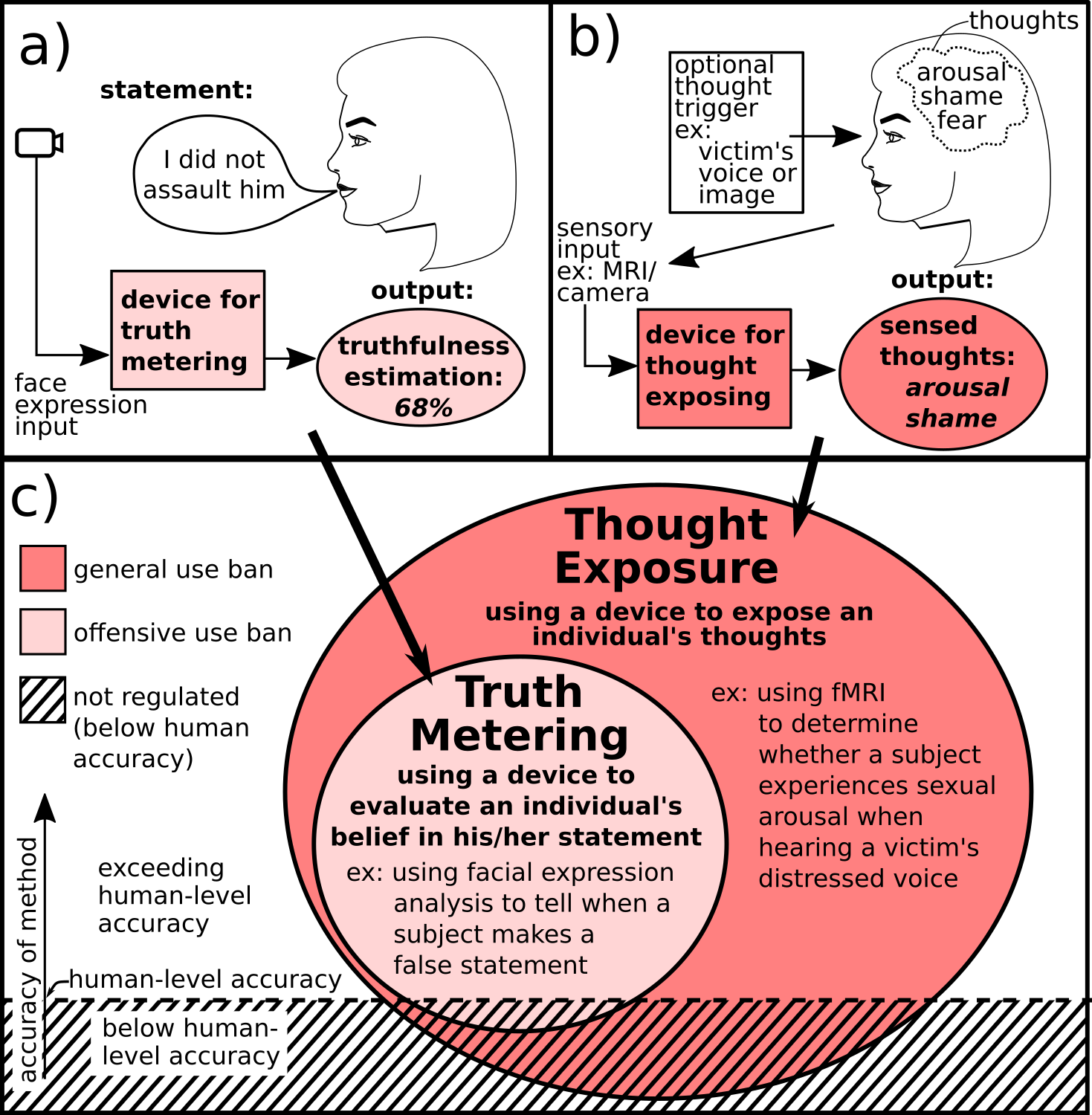}}
\caption{\textbf{Graphical Abstract of Proposed Mental Trespass Law.} Partitioning of lie detection technologies into \textit{truth metering} (a) and \textit{thought exposure} (b), and the proposed regulation of them (c).}
\label{fig_trespass}
\end{figure} 

Accurate thought exposing is defined as \textit{Use of a device to expose an individual's thoughts, without the individual's consent, with the device usage having an accuracy exceeding typical human performance.} Accurate thought exposure specifically includes instances of questioning a suspect without consent and accurately measuring the suspect's physiological response to the questions. As with the definition of truth metering, the definition of accurate thought exposure requires a level of accuracy which clearly surpasses human ability. A primary distinction of truth metering and accurate thought exposing, is that truth metering requires an overt/intentional statement by the individual regarding the issue being observed (Top portion of Fig. 2 illustrates this distinction). For example, in asking an individual what time it is, by evaluating whether they are being honest about the time involves only truth metering. However, if one then uses a system to gauge the level of anger in their voice, the technology has crossed the boundary into the realm of thought exposure because the overt response to the question being asked doesn't pertain to anger. Similarly, if an individual is talking out loud to others in public area on his/her own accord, and we evaluate the honesty of each of his/her overt statements, we are truth metering. However, if the individual's statements do not directly involve their emotions, and we determine that the individual feels high levels of arousal we are thought exposing (noting that a human observer would typically not be able to discern that information).


We concur with Greely and Illes that lie detection technologies and services must be regulated to prevent harm. Specifically, we believe that a federal Mental Trespass Act should be passed which: 1.) Provides a general ban on the use of ``accurate thought exposing" on an individual without the individual’s consent., 2.) Makes an exception to this ban for use of ``accurate truth metering" on individuals in a public space, as long as the particular usage would not be found offensive by a reasonable person, 3.) Updates the Employee Polygraph Protection Act to explicitly include ``accurate thought exposing" and ``accurate truth metering", even when such devices are noninvasive. 

\section{Discussion}
As much as these lie detection related technologies have the potential to infringe upon the civil liberties of the people in a malevolent way, there are a plethora of instances where the proper use of sophisticated, AI-driven sensing technologies and their associated algorithms can provide many benefits.

\subsection{Propagation of Fake News}
\begin{figure}[htbp]
\centerline{\includegraphics[width=1\linewidth]{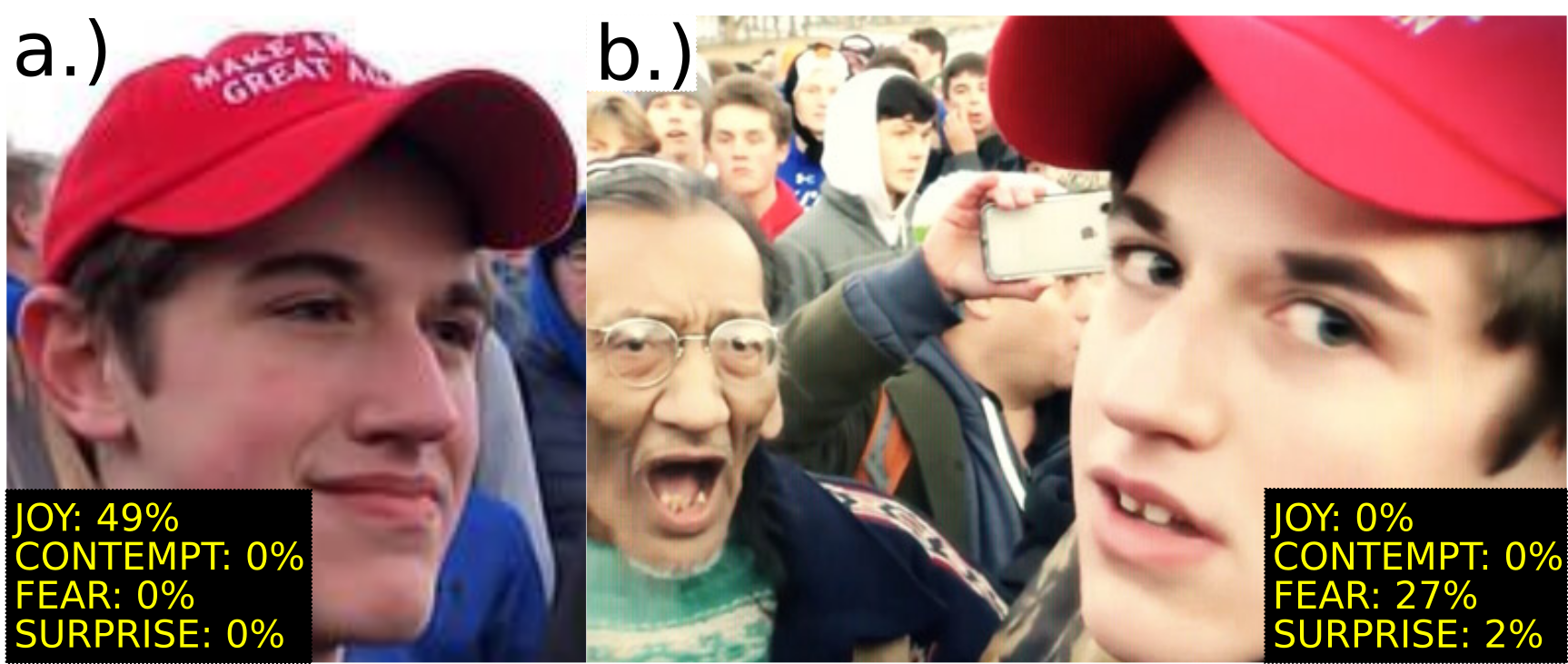}}
\caption[Infamous MAGA ``smirking teenager"]{\textbf{Infamous MAGA ``smirking teenager"} \cite{Noland2020} a) Nicholas Sandmann’s single frame expression taken out of context helped fuel a firestorm of rebuke even though automated facial expression analysis detects no contempt, b) analysis of another image from the event suggests that Sandmann was experiencing more fear and surprise.}
\label{fig_sandman}
\end{figure}

 Consider the infamous picture of the ``MAGA teen" Nicholas Sandmann (Figure \ref{fig_sandman}) that caused a recent social media and news firestorm as news commentators, actors, and numerous others, joined in for wave after wave of vilification towards Nicholas' Sandmann's alleged harassment of Native American Nathan Phillips \cite{grinberg2019}.

 Attention was drawn to Sandmann's expression, one CNN commentator tweeting ``Have you seen a more punchable face”, as others issued Sandman and his classmates death threats \cite{punchable}. Yet, an analysis of Sandmann’s face with facial expression analysis tools suggests that he was experiencing amusement rather than contempt. Indeed later reports validated Sandman’s assertions that he and his fellow classmates were not harassing the native American Nathan Phillips, but rather that the students were victims of harassment themselves \cite{realstory} (partially evidenced by Figure \ref{fig_sandman}b indicating that Sandman was feeling more afraid and caught off-guard). Had the news media used a noninvasive AI facial expression evaluation tool, this debacle (and other false news propagation like it) could have perhaps been avoided. 
 
 \subsection{Ensuring Fairness and Justice in the Courts}
 Although the Frye and Daubert Standards would certainly keep out accurate truth metering technologies from the courtrooms, there are approaches to circumvent the issues those standards posit to maximize the probability of obtaining justice in our Nation's court proceedings. We demonstrate these aspects using three components; An interview with a Judge, essential design primitives for developing accurate truth metering technologies and through one example, hope to show the ramifications for respecting an individual's cultural background.
\subsubsection{Interview with Judge}

In order to get an expert opinion on the potential impact advances in lie detection technologies could have on the courts, we interviewed standing County Judge Dennis Cohen of Livingston County, New York, who has twelve years experience on the bench. On the topic of emerging technologies in lie detection, Judge Cohen said \textit{``I think it is a big area of advancement in law, and could help us resolve cases and work through investigations ... just looking at what high resolution cameras have done for us with law shows that we can often identify the right culprit or prove that something happened or didn't happen."} Judge Cohen went on to say \textit{``Our whole society is changing because of technology. If it could be determined to be reliable ... then it could open up a whole new phase of things"}.  When asked about his opinion on relating the polygraph to these developing technologies threats and their associated threats of unreasonable searches, Cohen remarked \textit{``Polygraphs are voluntary. This [referring to these developing technologies] would also be a voluntary procedure as well, at least for the foreseeable future. Therefore it would not ever reach the bounds of an unreasonable search."} Here we see an important point brought up that when consent has been unquestionably obtained from an individual, usage of the polygraph or technologies to replace the polygraph never constitute an unreasonable search. However, the utility of such technologies designed in this way are vastly if not completely diminished due to their less than 100\% accuracy giving rise to the ``highly prejudicial" nature. Thus, it is prudent that in developing these technologies, that an entirely different approach be taken in their design primitives, development and deployment.

\subsubsection{Essential Design Primitives}
If accurate truth metering and/or thought exposure is used by law enforcement, it should be equally effective across all races and genders. Therefore, it is the responsibility of us and others who are researching and developing this technology to collect diverse data. We believe this could even be enforced by federal funding guidelines for those who are studying deception detection using Artificial Intelligence. In order to receive federal grants for this purpose, labs could be required to meet certain diversity standards in the data they collect and use in their deception detection algorithms. Additionally, the performance of said lie detection technologies should be standardized across all law enforcement entities.

Another relevant issue is how to maximize accuracy (as well as ability to deploy such devices in the court rooms) while preserving investigator autonomy. One solution proposed by Kleinberg et al., in their prediction framework for whether judges should jail or release criminal defendants, is to integrate the machine into the existing procedure, creating a human-machine symbiosis \cite{kleinberg2018human}. Instead of having the algorithm make all the decisions, the algorithm should give the people that are using it more information for them to make informed decisions themselves.

In this vein, it is our suggestion for researchers to create an output that is nuanced and detailed, rather than a binary 1 for "lying" and a 0 for "not lying." The lie detection device should detect and display indicators of deception when they appear. This fundamentally changes the role of the device. Instead of performing the evaluation based on an arbitrary decision boundary, it acts as a tool to assist people in doing the evaluation themselves. To interpret these more nuanced results, trained human operators should be employed. The use of such operators could even be required for the technology to be used. These operators should understand how to interpret the output and convey that information to investigators, while also understanding and conveying potential biases in certain questions as well as the potential for inaccuracy in the technology.

\subsubsection{Ramifications for Respecting Diverse Cultures}
The US can easily be viewed as a conglomeration of diversity given that most of the population can trace their family roots back to a family that immigrated to the US in the first place. This causes there to be a melting pot of different cultural backgrounds that inevitably find their way into the courtrooms. Challenged by how to integrate all these cultures successfully and fairly into the legal system, the AI-driven algorithms behind sensing technologies could provide valuable, novel solutions. 

Currently, the US legal framework does not support the wearing of masks in the courtroom. However, given the circumstances brought on by the COVID19 pandemic, this restriction has temporarily been lifted. This begs the question, should it ever have been a stipulation in the first place? Take for example a woman with a Muslim background who wishes to uphold her cultural traditions and wear a hijab during a court proceeding (e.g., she is called as a witness to bear testimony to the actions of another person). With our advanced sensing technology many options exist. First and foremost, the identity of the witness can be unquestionably established. This is perhaps the most important aspect to uphold. In the rarest of cases, let us for arguments sake assume that the wearing of a hijab interferes with one or two Jury members interpretation/perceived credibility of the witnesses testimony. In such a situation, human operators interpreting the results of the technology employed in the courtroom could be trained to identify this bias and address it. This is just one small example of where these advanced AI-driven sensing algorithms (which are also the backbone of facial recognition technology, creating deepfakes and filtering videos in general) can be used to treat every person that comes into the legal system with respect and fairness of the highest standard.

\subsection{Elaborations on Proposed Recommendations}
While dishonesty might frequently be harmful to people and society as a whole, we do believe that people have the right to exercise their ability to lie non-maliciously. Non-malicious lies are frequently altruistic or told by people to protect themselves or others, and allowing lie detection to remain unrestricted would prevent these kind of lies. We believe the harm caused by this would outweigh the benefit of allowing malicious lies to be detected, and therefore we believe that accurate thought exposing technologies should be regulated for the general public. Through establishing these regulations, we not only prevent potentially malicious uses, we offer further protections for the people against unreasonable searches of their mental sanctuaries. Recall in the case of Dow Chemical Co. v. United States, because the observation of the industrial complex was done through a high resolution camera and the general public had access to that technology, the court ruled that this did not constitute the bonds of an unreasonable search. With the public not having direct access to these emerging accurate thought exposing technologies, we thus prevent this legal precedent to be carried out in the future; enabling Government entities to take advantage of individuals in a variety of contexts.  It is our position that truth metering devices could remain available to the general public, as long as they were limited to uses that would be deemed non-offensive to a reasonable person. This would allow them to be used for lie detection in contexts such as navigating a foreign environment and dealing more fairly and justly with children. We formally take the stance that thought exposure systems must be regulated more strictly, as they can reveal more private information about a person (recall the unfortunate circumstances that led to the death of Tyler Clementi). We recommend that accurate thought exposing technologies be regulated for the general public (potentially by using a permit schema that is externally audited by multiple third parties relatively frequently), and that their unconsented use be codified as an illegal mental trespass. 

\section{Conclusion}
Accurate deception detection might not be available in the immediate future, although it is likely closer than most of us think. The technology's ambiguous legal status make it necessary to establish guidelines before it is fully developed and available. The introduction of AI-driven advanced sensing technologies for this task raises new concerns regarding privacy and consent due to their noninvasive nature. Defining the technologies precisely as ``accurate thought exposing" and ``accurate truth metering" technologies is essential for proposing legal doctrine that is as comprehensive and airtight as possible to safeguard civil liberties. Otherwise, potential loopholes could emerge in the future causing harm to society and bypassing the intentions of the law and the protections that it offers (as is the likely case currently with No Lie fMRI and the EPPA Act). Emerging lie detection technology will be a powerful tool, benefiting the criminal justice system, the medical community, and many others. In order to utilize it to its fullest potential, however, it must be developed and used responsibly with the necessary restrictions - or it may end up doing more harm than good. Rather than nurture and help his creation acclimatize to society, Dr. Frankenstein rejected him. If we continue to ignore the legal status of coming AI technologies, rather than nurture and regulate them, we too may end up with a Frankenstein monster as Larson proclaimed.

\bibliographystyle{aaai.sty}
\bibliography{01_main.bbl}

\begin{thebibliography}{}

\bibitem[\protect\citeauthoryear{Adkins}{2001}]{adkins2001supreme}
Adkins, D.
\newblock 2001.
\newblock The supreme court announces a fourth amendment general public use
  standard for emerging technologies but fails to define it: Kyllo v. united
  states.
\newblock {\em U. Dayton L. Rev.} 27:245.

\bibitem[\protect\citeauthoryear{Alder}{2007}]{alder2007lie}
Alder, K.
\newblock 2007.
\newblock {\em The lie detectors: The history of an American obsession}.
\newblock Simon and Schuster.

\bibitem[\protect\citeauthoryear{Amar and Lettow}{1995}]{amar1995fifth}
Amar, A.~R., and Lettow, R.~B.
\newblock 1995.
\newblock Fifth amendment first principles: The self-incrimination clause.
\newblock {\em Michigan Law Review} 93(5):857--928.

\bibitem[\protect\citeauthoryear{Amar}{1994}]{amar1994fourth}
Amar, A.~R.
\newblock 1994.
\newblock Fourth amendment first principles.
\newblock {\em Harvard Law Review} 107(4):757--819.

\bibitem[\protect\citeauthoryear{Amsel}{2019}]{amsel2019planting}
Amsel, T.~T.
\newblock 2019.
\newblock Planting the seeds of polygraph's practice a brief historical review.
\newblock {\em European Polygraph} 13(3):141--154.

\bibitem[\protect\citeauthoryear{BBC}{2020}]{bbc2020}
BBC, T.
\newblock 2020.
\newblock The uighurs and the chinese state: A long history of discord.
\newblock BBC News.
\newblock Retrieved November 15, 2020 from
  \url{https://www.bbc.com/news/world-asia-china-22278037}.

\bibitem[\protect\citeauthoryear{Bekiares}{2005}]{bekiares2005constitutional}
Bekiares, J.~A.
\newblock 2005.
\newblock Constitutional law: Ratifying suspicionless canine sniffs: Dog days
  on the highways-illinois v. caballes.
\newblock {\em Fla. L. Rev.} 57:963.

\bibitem[\protect\citeauthoryear{Bond~Jr and DePaulo}{2006}]{bond2006accuracy}
Bond~Jr, C.~F., and DePaulo, B.~M.
\newblock 2006.
\newblock Accuracy of deception judgments.
\newblock {\em Personality and social psychology Review} 10(3):214--234.

\bibitem[\protect\citeauthoryear{Bond~Jr and
  DePaulo}{2008}]{bond2008individual}
Bond~Jr, C.~F., and DePaulo, B.~M.
\newblock 2008.
\newblock Individual differences in judging deception: Accuracy and bias.
\newblock {\em Psychological bulletin} 134(4):477.

\bibitem[\protect\citeauthoryear{Bradley and
  Klohn}{1987}]{bradley1987machiavellianism}
Bradley, M., and Klohn, K.
\newblock 1987.
\newblock Machiavellianism, the control question test and the detection of
  deception.
\newblock {\em Perceptual and Motor Skills} 64(3):747--757.

\bibitem[\protect\citeauthoryear{Brey}{2004}]{brey2004ethical}
Brey, P.
\newblock 2004.
\newblock Ethical aspects of facial recognition systems in public places.
\newblock {\em Journal of information, communication and ethics in society}.

\bibitem[\protect\citeauthoryear{Bureau~of Labor~Stat.}{2018}]{BLS}
Bureau~of Labor~Stat., B.
\newblock 2018.
\newblock Cpi inflation calculator.
\newblock Retrieved August 7, 2018 from
  \url{https://www.bls.gov/data/inflation_calculator.htm}.

\bibitem[\protect\citeauthoryear{Cauthen}{2017}]{cauthen2017fifth}
Cauthen, R.~H.
\newblock 2017.
\newblock The fifth amendment and compelling unencrypted data, encryption
  codes, and passwords.
\newblock {\em Am. J. Trial Advoc.} 41:119.

\bibitem[\protect\citeauthoryear{Congress}{2019}]{oversight2019}
Congress, U.
\newblock 2019.
\newblock Facial recognition technology: Part i its impact on our civil rights
  and liberties.
\newblock 116th Congress, Ser. No. 116027, Committee on Oversight and Reform,
  Retrieved January 30, 2021 from
  \url{https://docs.house.gov/meetings/GO/GO00/20190522/109521/HHRG-116-GO00-Transcript-20190522.pdf}.

\bibitem[\protect\citeauthoryear{Cunis}{1988}]{cunis1988california}
Cunis, D.~W.
\newblock 1988.
\newblock California v. greenwood: Discarding the traditional approach to the
  search and seizure of garbage.
\newblock {\em Cath. UL Rev.} 38:543.

\bibitem[\protect\citeauthoryear{Darwin}{1873}]{Darwin1873}
Darwin, C.
\newblock 1873.
\newblock {\em The Expression of the Emotions in Man and Animals}.
\newblock D. Appleton.

\bibitem[\protect\citeauthoryear{Dery~III}{2005}]{dery2005let}
Dery~III, G.~M.
\newblock 2005.
\newblock Who let the dogs out-the supreme court did in illinois v. caballes by
  placing absolute faith in canine sniffs.
\newblock {\em Rutgers L. Rev.} 58:377.

\bibitem[\protect\citeauthoryear{Falcone}{1986}]{falcone1986california}
Falcone, R.
\newblock 1986.
\newblock California v. ciraolo: The demise of private property.
\newblock {\em La. L. Rev.} 47:1365.

\bibitem[\protect\citeauthoryear{Farah \bgroup et al\mbox.\egroup
  }{2014}]{farah2014functional}
Farah, M.~J.; Hutchinson, J.~B.; Phelps, E.~A.; and Wagner, A.~D.
\newblock 2014.
\newblock Functional mri-based lie detection: scientific and societal
  challenges.
\newblock {\em Nature Reviews Neuroscience} 15(2):123--131.

\bibitem[\protect\citeauthoryear{Fong, Dey, and
  Chaki}{2020}]{fong2020artificial}
Fong, S.~J.; Dey, N.; and Chaki, J.
\newblock 2020.
\newblock Artificial intelligence for coronavirus outbreak.

\bibitem[\protect\citeauthoryear{Froh}{2002}]{froh2002rethinking}
Froh, A.~S.
\newblock 2002.
\newblock Rethinking canine sniffs: The impact of kyllo v. united states.
\newblock {\em Seattle UL Rev.} 26:337.

\bibitem[\protect\citeauthoryear{Goldzband}{1990}]{goldzband1990polygraph}
Goldzband, M.~G.
\newblock 1990.
\newblock The polygraph and psychiatrists.
\newblock {\em Journal of Forensic Science} 35(2):391--402.

\bibitem[\protect\citeauthoryear{Goodfellow \bgroup et al\mbox.\egroup
  }{2014}]{goodfellow2014generative}
Goodfellow, I.; Pouget-Abadie, J.; Mirza, M.; Xu, B.; Warde-Farley, D.; Ozair,
  S.; Courville, A.; and Bengio, Y.
\newblock 2014.
\newblock Generative adversarial nets.
\newblock In {\em Advances in neural information processing systems},
  2672--2680.

\bibitem[\protect\citeauthoryear{Greely and
  Illes}{2007}]{greely2007neuroscience}
Greely, H.~T., and Illes, J.
\newblock 2007.
\newblock Neuroscience-based lie detection: The urgent need for regulation.
\newblock {\em American Journal of Law \& Medicine} 33(2-3):377--431.

\bibitem[\protect\citeauthoryear{Grinberg}{2019}]{grinberg2019}
Grinberg, E.
\newblock 2019.
\newblock Videos show a collision of 3 groups that spawned a fiery political
  moment.
\newblock January 30, 2021 from
  \url{https://www.nytimes.com/2019/01/22/us/covington-catholic-washington-videos.html}.

\bibitem[\protect\citeauthoryear{Grubin and Madsen}{2005}]{grubin2005lie}
Grubin, D., and Madsen, L.
\newblock 2005.
\newblock Lie detection and the polygraph: A historical review.
\newblock {\em The Journal of Forensic Psychiatry \& Psychology}
  16(2):357--369.

\bibitem[\protect\citeauthoryear{Herdrich}{1988}]{herdrich1988california}
Herdrich, M.~A.
\newblock 1988.
\newblock California v. greenwood: The trashing of privacy.
\newblock {\em Am. UL Rev.} 38:993.

\bibitem[\protect\citeauthoryear{Hochreutiner}{2020}]{Hochreutiner2020}
Hochreutiner, C.
\newblock 2020.
\newblock How smarter ai™-powered cameras can mitigate the spread of wuhan
  novel coronavirus (covid-19), and what we’ve learned from the sars outbreak
  17 years prior.
\newblock anyconnect.com.
\newblock Retrieved August 8, 2018 from
  \url{https://anyconnect.com/blog/smart-thermal-cameras-wuhan-coronavirus}.

\bibitem[\protect\citeauthoryear{Hosch}{2018}]{Hosch2018}
Hosch, W.~L.
\newblock 2018.
\newblock Supercomputer.
\newblock Encyclopedia Britannica.
\newblock Retrieved August 7, 2018 from
  \url{https://www.britannica.com/technology/supercomputer}.

\bibitem[\protect\citeauthoryear{Hruska}{2016}]{Hruska2016}
Hruska, J.
\newblock 2016.
\newblock Nvidia titan x offers incredible performance, at a significant cost.
\newblock ExtremeTech.
\newblock Retrieved August 7, 2018 from
  \url{https://www.extremetech.com/gaming/232924-nvidia-titan-x-offers-incredible/performance-at-a-significant-cost}.

\bibitem[\protect\citeauthoryear{Inbau}{1998}]{inbau1998self}
Inbau, F.~E.
\newblock 1998.
\newblock Self-incrimination--what can an accused person be compelled to do?
\newblock {\em J. Crim. L. \& Criminology} 89:1329.

\bibitem[\protect\citeauthoryear{Jensen}{2002}]{jensen2002frye}
Jensen, P.~J.
\newblock 2002.
\newblock Frye versus daubert: Practically the same.
\newblock {\em Minn. Law Review} 87:1579.

\bibitem[\protect\citeauthoryear{Joyce}{1982}]{joyce1982epa}
Joyce, T.~J.
\newblock 1982.
\newblock The epa's use of aerial photography violates the fourth amendment:
  Dow chemical co. v. united states.
\newblock {\em CoNN. L. REv.} 15:327.

\bibitem[\protect\citeauthoryear{Kittay}{2006}]{kittay2006admissibility}
Kittay, L.
\newblock 2006.
\newblock Admissibility of fmri lie detection-the cultural bias against mind
  reading devices.
\newblock {\em Brook. L. Rev.} 72:1351.

\bibitem[\protect\citeauthoryear{Kleinberg \bgroup et al\mbox.\egroup
  }{2018}]{kleinberg2018human}
Kleinberg, J.; Lakkaraju, H.; Leskovec, J.; Ludwig, J.; and Mullainathan, S.
\newblock 2018.
\newblock Human decisions and machine predictions.
\newblock {\em The quarterly journal of economics} 133(1):237--293.

\bibitem[\protect\citeauthoryear{Krizhevsky, Sutskever, and
  Hinton}{2012}]{krizhevsky2012imagenet}
Krizhevsky, A.; Sutskever, I.; and Hinton, G.~E.
\newblock 2012.
\newblock Imagenet classification with deep convolutional neural networks.
\newblock In {\em Advances in neural information processing systems},
  1097--1105.

\bibitem[\protect\citeauthoryear{Langleben and
  Moriarty}{2013}]{langleben2013using}
Langleben, D.~D., and Moriarty, J.~C.
\newblock 2013.
\newblock Using brain imaging for lie detection: Where science, law, and policy
  collide.
\newblock {\em Psychology, Public Policy, and Law} 19(2):222.

\bibitem[\protect\citeauthoryear{Lee \bgroup et al\mbox.\egroup
  }{2018}]{lee2018design}
Lee, S.; Park, Y.-D.; Suh, Y.-J.; and Jeon, S.
\newblock 2018.
\newblock Design and implementation of monitoring system for breathing and
  heart rate pattern using wifi signals.
\newblock In {\em 2018 15th IEEE Annual Consumer Communications \& Networking
  Conference (CCNC)},  1--7.
\newblock IEEE.

\bibitem[\protect\citeauthoryear{McCall}{1996}]{mccall1996misconceptions}
McCall, J.~R.
\newblock 1996.
\newblock Misconceptions and reevaluation-polygraph admissibility after rock
  and daubert.
\newblock {\em U. Ill. L. Rev.}  363.

\bibitem[\protect\citeauthoryear{Moreno}{2009}]{moreno2009future}
Moreno, J.~A.
\newblock 2009.
\newblock The future of neuroimaged lie detection and the law.
\newblock {\em Akron L. Rev.} 42:717.

\bibitem[\protect\citeauthoryear{Myers and
  Arbuthnot}{1997}]{myers1997polygraph}
Myers, B., and Arbuthnot, J.
\newblock 1997.
\newblock Polygraph testimony and juror judgments: A comparison of the guilty
  knowledge test and the control question test 1.
\newblock {\em Journal of Applied Social Psychology} 27(16):1421--1437.

\bibitem[\protect\citeauthoryear{Noland}{2020}]{Noland2020}
Noland, K.
\newblock 2020.
\newblock January 2019 lincoln memorial confrontation.
\newblock youtube \& wikipedia.
\newblock Retrieved November 15, 2020 from
  \url{http://img.sina.com/translate/781/w500h281/20190121/7dYk-hryfqhk3793301.jpg}.

\bibitem[\protect\citeauthoryear{NVIDIA}{2018}]{NVIDIA2018}
NVIDIA.
\newblock 2018.
\newblock Geforce gtx titan x.
\newblock Retrieved August 7, 2018 from
  \url{https://www.geforce.com/hardware/desktop-gpus/geforce-gtx-titan-x/specifications}.

\bibitem[\protect\citeauthoryear{Peterson}{1989}]{peterson1989bud}
Peterson, E.~J.
\newblock 1989.
\newblock Bud lent and doc cambell: Two esteemed justices of the oregon supreme
  court.
\newblock {\em Willamette L. Rev.} 25:243.

\bibitem[\protect\citeauthoryear{Pilkington}{2010}]{Pilkington2010}
Pilkington, E.
\newblock 2010.
\newblock Tyler clementi, student outed as gay on internet, jumps to his death.
\newblock The Guardian.
\newblock Retrieved August 8, 2018 from
  \url{https://www.theguardian.com/world/2010/sep/30/tyler-clementi-gay-student-suicide}.

\bibitem[\protect\citeauthoryear{Pitt}{2012}]{pitt2012design}
Pitt, J.
\newblock 2012.
\newblock Design contractualism for pervasive/affective computing.
\newblock {\em IEEE Technology and Society Magazine} 31(4):22--29.

\bibitem[\protect\citeauthoryear{Raji \bgroup et al\mbox.\egroup
  }{2020}]{raji2020saving}
Raji, I.~D.; Gebru, T.; Mitchell, M.; Buolamwini, J.; Lee, J.; and Denton, E.
\newblock 2020.
\newblock Saving face: Investigating the ethical concerns of facial recognition
  auditing.
\newblock In {\em Proceedings of the AAAI/ACM Conference on AI, Ethics, and
  Society},  145--151.

\bibitem[\protect\citeauthoryear{Raskin and Honts}{2002}]{raskin2002comparison}
Raskin, D.~C., and Honts, C.~R.
\newblock 2002.
\newblock The comparison question test.

\bibitem[\protect\citeauthoryear{Reynolds and
  Picard}{2004}]{reynolds2004affective}
Reynolds, C., and Picard, R.
\newblock 2004.
\newblock Affective sensors, privacy, and ethical contracts.
\newblock In {\em CHI'04 extended abstracts on Human factors in computing
  systems},  1103--1106.

\bibitem[\protect\citeauthoryear{Reynolds and
  Picard}{2005}]{reynolds2005evaluation}
Reynolds, C., and Picard, R.~W.
\newblock 2005.
\newblock Evaluation of affective computing systems from a dimensional
  metaethical position.
\newblock In {\em First Augmented Cognition International Conference, Las
  Vegas, NV}.

\bibitem[\protect\citeauthoryear{Richardson}{2020}]{punchable}
Richardson, V.
\newblock 2020.
\newblock Ex-cnn host 'likely' to be sued over now-deleted 'punchable face'
  tweet: Sandmann attorney.
\newblock The Washington Times.
\newblock Retrieved November 10, 2030 from
  \url{https://www.washingtontimes.com/news/2020/jan/13/reza-aslan-likely-be-sued-over-now-/deleted-punchab/}.

\bibitem[\protect\citeauthoryear{Roth}{1995}]{roth1995mesopotamian}
Roth, M.~T.
\newblock 1995.
\newblock Mesopotamian legal traditions and the laws of hammurabi.
\newblock {\em Chi.-Kent L. Rev.} 71:13.

\bibitem[\protect\citeauthoryear{Rubin}{2003}]{rubin2003square}
Rubin, P.~J.
\newblock 2003.
\newblock Square pegs and round holes: Substantive due process, procedural due
  process, and the bill of rights.
\newblock {\em Columbia Law Review}  833--892.

\bibitem[\protect\citeauthoryear{Ruzi}{1988}]{ruzi1988reviving}
Ruzi, S.~H.
\newblock 1988.
\newblock Reviving trespass-based search analysis under the open view doctrine:
  Dow chemical co. v. united states.
\newblock {\em NYUL Rev.} 63:191.

\bibitem[\protect\citeauthoryear{Seamon}{2001}]{seamon2001kyllo}
Seamon, R.~H.
\newblock 2001.
\newblock Kyllo v. united states and the partial ascendance of justice scalia's
  fourth amendment.
\newblock {\em Wash. ULQ} 79:1013.

\bibitem[\protect\citeauthoryear{Sen \bgroup et al\mbox.\egroup
  }{2018}]{sen2018automated}
Sen, T.; Hasan, M.~K.; Teicher, Z.; and Hoque, M.~E.
\newblock 2018.
\newblock Automated dyadic data recorder (addr) framework and analysis of
  facial cues in deceptive communication.
\newblock {\em Proceedings of the ACM on Interactive, Mobile, Wearable and
  Ubiquitous Technologies} 1(4):1--22.

\bibitem[\protect\citeauthoryear{Simmons}{2005}]{simmons2005two}
Simmons, R.
\newblock 2005.
\newblock The two unanswered questions of illinois v. caballes: How to make the
  world safe for binary searches.
\newblock {\em Tul. L. Rev.} 80:411.

\bibitem[\protect\citeauthoryear{Simpson}{1989}]{simpson1989california}
Simpson, D.
\newblock 1989.
\newblock California v. greenwood: The pruning of the fourth amendment.
\newblock {\em Loy. L. REv.} 35:549.

\bibitem[\protect\citeauthoryear{Soave}{2020}]{realstory}
Soave, R.
\newblock 2020.
\newblock A year ago, the media mangled the covington catholic story. what
  happened next was even worse.
\newblock
  \url{https://reason.com/2020/01/21/covington-catholic-media-nick-sandmann-/lincoln-memorial/}.
\newblock Accessed: 2020-08-21.

\bibitem[\protect\citeauthoryear{Spicer}{2000}]{Spicer2000}
Spicer, D.
\newblock 2000.
\newblock Control data 6600: The supercomputer arrives.
\newblock Retrieved August 7, 2018 from
  \url{http://www.drdobbs.com/control-data-6600-the-supercomputer-arr/i/184404102}.

\bibitem[\protect\citeauthoryear{Tennison and
  Moreno}{2012}]{tennison2012neuroscience}
Tennison, M.~N., and Moreno, J.~D.
\newblock 2012.
\newblock Neuroscience, ethics, and national security: the state of the art.
\newblock {\em PLoS Biol} 10(3):e1001289.

\bibitem[\protect\citeauthoryear{Thompson}{2007}]{thompson2007brave}
Thompson, S.~K.
\newblock 2007.
\newblock A brave new world of interrogation jurisprudence?
\newblock {\em American journal of law \& medicine} 33(2-3):341--357.

\bibitem[\protect\citeauthoryear{Trovillo}{1938}]{trovillo1938history}
Trovillo, P.~V.
\newblock 1938.
\newblock History of lie detection.
\newblock {\em Am. Inst. Crim. L. \& Criminology} 29:848.

\bibitem[\protect\citeauthoryear{Trovillo}{1939}]{Trovillo1939}
Trovillo, P.~V.
\newblock 1939.
\newblock History of lie detection.
\newblock {\em Journal of Criminal Law and Criminology} 29.

\bibitem[\protect\citeauthoryear{US~Court~of Appeals~D.C.}{1923}]{Frye1923}
US~Court~of Appeals~D.C., C.
\newblock 1923.
\newblock Frye v. united states.
\newblock 293 F. 1013 (D.C. Cir. 1923).

\bibitem[\protect\citeauthoryear{Wachtel}{2013}]{wachtel2013give}
Wachtel, M.
\newblock 2013.
\newblock Give me your password because congress can say so: An analysis of
  fifth amendment protection afforded individuals regarding compelled
  production of encrypted data and possible solutions to the problem of getting
  data from someone's mind.
\newblock {\em Pitt. J. Tech. L. \& Pol'y} 14:44.

\bibitem[\protect\citeauthoryear{Wang and Kosinski}{2018}]{wang2018deep}
Wang, Y., and Kosinski, M.
\newblock 2018.
\newblock Deep neural networks are more accurate than humans at detecting
  sexual orientation from facial images.
\newblock {\em Journal of personality and social psychology} 114(2):246.

\bibitem[\protect\citeauthoryear{Wang, Pun, and
  Chanel}{2018}]{wang2018comparative}
Wang, C.; Pun, T.; and Chanel, G.
\newblock 2018.
\newblock A comparative survey of methods for remote heart rate detection from
  frontal face videos.
\newblock {\em Frontiers in bioengineering and biotechnology} 6:33.

\bibitem[\protect\citeauthoryear{Weintraub}{1956}]{weintraub1956voice}
Weintraub, R.~J.
\newblock 1956.
\newblock Voice identification, writing exemplars and the privilege against
  self-incrimination.
\newblock {\em Vand. L. Rev.} 10:485.

\bibitem[\protect\citeauthoryear{Wolpe, Foster, and
  Langleben}{2005}]{wolpe2005emerging}
Wolpe, P.~R.; Foster, K.~R.; and Langleben, D.~D.
\newblock 2005.
\newblock Emerging neurotechnologies for lie-detection: promises and perils.
\newblock {\em The American Journal of Bioethics} 5(2):39--49.

\end{thebibliography}

\end{document}